\documentclass[11pt]{article}
\usepackage{abstract}
\usepackage{amsbsy}
\usepackage{amscd}
\usepackage{amsfonts}
\usepackage{amsmath}
\usepackage{amssymb}
\usepackage{amsthm}
\usepackage{bm}
\usepackage{calligra}
\usepackage{cite}
\usepackage{epsfig}
\usepackage{epstopdf}
\usepackage{float}
\usepackage{graphicx}
\usepackage{overpic}
\usepackage[margin=0.5in]{geometry}
\usepackage{mathrsfs}
\usepackage{mathtools}
\usepackage{makecell}
\usepackage{multirow}
\usepackage{xcolor}
\usepackage[T1]{fontenc}
\usepackage{psfrag}
\usepackage{subfigure}
\usepackage{booktabs}
\usepackage{listings}
\usepackage{longtable}
\usepackage{latexsym}
\usepackage{arydshln}
\usepackage{enumerate}
\usepackage{verbatim}
\usepackage{caption}
\allowdisplaybreaks[4]

\date{}
\title{Dynamical System and Statefinder Analysis of Cosmological Models in $f(T, B)$ Gravity}

\author{Jianwen~Liu, Fabao~Gao*, Aqeela~Razzaq\\
	\it\footnotesize School of Mathematical Science, Yangzhou University, Yangzhou 225002, P.R.\,China\\
\footnotesize {}*Correspondence: gaofabao@sina.com or fbgao@yzu.edu.cn}

\newtheorem {theorem*}{Theorem}

\numberwithin{equation}{section}

\usepackage[colorlinks,
            CJKbookmarks=true,
            linkcolor=blue,
            anchorcolor=red,
             urlcolor=blue,
            citecolor=blue
            ]{hyperref}

\begin{document}

\maketitle
\begin{abstract}
This study systematically investigates the cosmological dynamics of two well-motivated functional forms in $f(T,B)$ gravity within a flat Friedmann-Lema\^{i}tre-Robertson-Walker (FLRW) universe. Here $T$ denotes the torsion scalar and $B$ the boundary term, with the special choice $f(T,B) = - T + B$ recovering General Relativity. We focus on a multiplicative power-law model $f(T,B) = c_1 T^\alpha B^\beta$ and an additive mixed power-law model $f(T,B) = c_2 T^\alpha + c_3 B^\beta$. Using dynamical system techniques, we construct autonomous systems and identify de Sitter attractors that naturally explain late-time cosmic acceleration. Analytical stability conditions for these fixed points are derived, and numerical simulations reveal characteristic evolutionary patterns, such as spiral trajectories and damped oscillations in the additive mixed power-law model. Furthermore, statefinder diagnostics are applied to quantitatively distinguish these models from the standard $\Lambda$CDM paradigm and other dark energy scenarios. The results indicate that $f(T,B)$ gravity offers a theoretically consistent and observationally distinguishable geometric framework for explaining cosmic acceleration, presenting a compelling alternative to conventional dark energy models.
%
\vskip 0.2cm
%
\end{abstract}

\section{Introduction} \label{section-1}
A major puzzle in modern physics is the universe's late-time acceleration \cite{Riess 1998, Perlmutter 1999, Peebles 2003} and the observed mass discrepancy in large-scale structures \cite{Persic 1996}, whose dynamics cannot be explained by visible matter \cite{Bertone 2018}. These phenomena are often attributed to dark energy and dark matter -- hypothetical components that influence cosmic evolution despite having no direct electromagnetic signatures.

Dark energy presents a particular enigma due to its repulsive gravitational effect, which counteracts conventional gravitational attraction. Two major theoretical approaches have been developed to address these cosmological puzzles. The first retains the geometric framework of General Relativity (GR) while introducing new matter components, such as scalar fields with negative pressure. The second approach fundamentally modifies gravitational theory itself, either by extending the geometry underlying Einstein's field equations or by reinterpreting how matter influences spacetime curvature. Such modified gravity theories generally generalize the Einstein-Hilbert action via various geometric extensions, giving rise to several well-established frameworks. These include curvature-based modifications like $f(R)$ \cite{Buchdahl 1970, Felice 2010a, Sotiriou 2010} and $f(G)$ gravity \cite{Nojiri 2005, Felice 2009}, higher-order polynomial extensions such as cubic gravity \cite{Bueno 2016, Erices 2019, Asimakis 2024}, topological invariants in Lovelock gravity \cite{Concha 2017}, and scalar-tensor couplings exemplified by Horndeski's theory \cite{Kobayashi 2019} and its Galileon generalizations \cite{Nicolis 2009, Felice 2010b}. This dichotomy underscores the central debate in modern cosmology: whether dark energy originates from new material constituents or from an extension of gravitational theory itself.

An alternative formulation of gravity, dynamically equivalent to GR at the level of field equations, employs torsion rather than curvature as the fundamental geometric descriptor. This approach, known as the Teleparallel Equivalent of General Relativity (TEGR) \cite{Maluf 2013, Bahamonde 2015}, describes gravity through a torsion-based geometry where gravitational interactions result from the parallel transport of tetrad fields. In TEGR, the gravitational field is characterized by the torsion tensor with dynamics governed by the torsion scalar $T$, a quadratic contraction of the torsion tensor. A further equivalent representation of Einstein’s theory can be formulated in a flat, torsion-free geometry where gravitation is fully encoded in the non-metricity tensor $Q_{\alpha \mu \nu}=\nabla_\alpha g_{\mu\nu}$. This framework, known as symmetric teleparallel gravity, leads to $f(Q)$ gravity \cite{Jimenez 2018, Jarv 2018, Jimenez 2020}; further details on this theory and its extensions can be found in \cite{Mandal 2020, Lazkoz 2019, Heisenberg 2024, Gadbail 2024, Gadbail 2023}. Although TEGR reproduces the predictions of GR, it offers a distinct geometric interpretation \cite{Geng 2012, Krssak 2019, Bahamonde 2023}, casting gravity as a manifestation of spacetime torsion rather than curvature. Nevertheless, like GR, TEGR alone does not resolve large-scale cosmological issues such as dark energy or inflation. To address these limitations, modified teleparallel theories, collectively referred to as $f(T)$ gravity, have been developed \cite{Ferraro 2007, Ferraro 2008, Bengochea 2009}, generalizing the Lagrangian to arbitrary functions of $T$. As an extension of TEGR, $f(T)$ gravity opens new pathways for explaining cosmic acceleration and large-scale structure formation \cite{Wu 2010, Capozziello 2018, Capozziello 2022, Kavya 2024}. Still, whether $f(T)$ theories can outperform GR on both theoretical and observational grounds remains an open question, warranting further detailed investigation.

To construct a complete teleparallel analog of $f(R)$ gravity, the $f(T, B)$ extension plays an essential role \cite{Bahamonde 2015, Wright 2016, Bahamonde 2017}. In this framework, the torsion scalar $T$ and the boundary term $B$ respectively capture the second- and fourth-order derivative contributions present in $f(R)$ gravity. Notably, $f(T, B)$ naturally incorporates $f(R)$ gravity as the specific case $f(-T+B)$, while allowing for a wider range of gravitational Lagrangians.

The $f(T, B)$ framework has been widely explored across diverse phenomenological contexts. Gravitational wave studies in this theory indicate luminal propagation speeds and the presence of polarization modes beyond the standard transverse-traceless ones of GR \cite{Farrugia 2018, Capozziello 2020}. Solar-system tests further confirm the viability of many $f(T, B)$ models, showing agreement with high-precision astronomical measurements \cite{Farrugia 2020}. In cosmology, such models offer promising mechanisms to alleviate the Hubble tension \cite{Rivera 2020, Franco 2021, Briffa 2023}. Recent theoretical developments include: (i) a rigorous establishment of the correspondence between $f(T, B)$ and $f(R)$ gravity \cite{Bahamonde 2015, Wright 2016}, along with thermodynamic and cosmological reconstruction studies \cite{Bahamonde 2018}; (ii) exact and perturbed black hole solutions \cite{Bahamonde 2022}, extending the theory's applicability to compact objects; and (iii) comprehensive analyses of background expansion and linear perturbation growth \cite{Paliathanasis 2017, Franco 2020, Caruana 2020, Samaddar 2023, Kadam 2023}.

This study investigates the cosmological dynamics of $f(T,B)$ gravity within an isotropic, homogeneous FLRW universe using the dynamical system approach. The strong nonlinearity of the field equations renders exact solutions intractable and obstructs direct observational tests, owing to the complex coupling among terms. To address this, the dynamical system method \cite{Wainwright 1997, Coley 2003,Bahamonde 2018}, which has been used to investigate cosmological models in various modified gravity, including Ho\v{r}ava-Lifshitz gravity \cite{Gao 2019, Gao 2020, Gao 2021, Gao 2022}, $f(R)$ gravity \cite{MacDevette 2025,Liu 2025}, $f(R, T)$ gravity \cite{Liu 2022,Goncalves 2024}, Gauss-Bonnet gravity \cite{Singh 2025}, Einstein cubic gravity \cite{Mandal 2025}, and other cosmological scenarios \cite{Papagiannopoulos 2020,Singh 2023}.  This method reformulates the cosmological equations into an autonomous system, enabling the examination of critical points, phase-space trajectories, and stability, yielding global dynamical insight (including attractors and transient states) without relying on exact solutions. Such a global perspective is essential for comparing theories with observations, as it reveals the full dynamical landscape rather than individual solutions. Recently, using the dynamical system method, Kritpetch \textit{et al}. \cite{Kritpetch 2025} clarified the interaction mechanism between dark sector components in dark energy models incorporating both quintessence and phantom fields via a switching parameter. In parallel, Halder \textit{et al}. \cite{Halder 2025} identified new stable accelerating scaling attractors within interacting phantom dark energy frameworks. These attractors offer a potential mechanism for alleviating the cosmic coincidence problem.

This study is structured as follows. Section \ref{section-2} provides a brief review of teleparallel gravity and its extensions. In Section \ref{section-3}, we derive the cosmological dynamical system for the $f(T,B)$ model. Section \ref{section-4} presents a dynamical analysis of the two considered models, and Section \ref{section-5} examines their statefinder diagnostics. Finally, Section \ref{section-6} summarizes the key findings and offers a concluding discussion.

\section{Teleparallel gravity and its extension $f(T, B)$ gravity} \label{section-2}
This section provides a brief review of Teleparallel Gravity (TG) and its extension to $f(T, B)$ gravity. GR describes gravity using the Levi-Civita connection $\Gamma^a_{\mu\nu}$, which is characterized by non-zero curvature, zero torsion, and metric compatibility. In contrast, Teleparallel Gravity adopts the Weitzenb\"ock connection $W_\mu{}^a{}_\nu$ -- a curvature-free, metric-compatible connection that captures gravitational effects entirely through torsion \cite{Cai 2016}. This shift in geometric foundation has profound implications: while GR and its modifications employ the Riemann tensor to measure spacetime curvature, the Riemann tensor vanishes identically in TG due to the flatness of the Weitzenb\"ock connection. As a result, TG necessitates a reconstruction of gravitational quantities from the ground up, offering novel theoretical possibilities while preserving dynamical equivalence with GR at the level of field equations.

In the TG framework, the fundamental dynamical variables are the tetrad fields (vierbeins) $e^a_\mu$, which form an orthonormal basis for the tangent space at each spacetime point $x^\mu$. The tetrads $e^a_\mu$ and their inverse fields $E^\mu_a$ satisfy the orthonormality conditions
\begin{equation} \label{equation-2.1}
\begin{aligned}
e^m_\mu E^\mu_n = \delta^m_n, \quad
e^m_\mu E^\nu_m = \delta^\nu_\mu,
\end{aligned}
\end{equation}
where Latin indices $(m, n)$ refer to coordinates in the tangent space, and Greek indices $(\mu, \nu)$ denote spacetime coordinates. The metric tensor $g_{\mu\nu}$ is reconstructed from the tetrad fields via the relation
\begin{equation*}
	g_{\mu\nu}=e^a_\mu e^b_\nu \eta_{ab},
\end{equation*}
where $\eta_{ab}$ is the Minkowski metric on the tangent space. The Weitzenb\"ock connection is defined as
\cite{Bahamonde 2015}
\begin{equation*}
	{W_\mu}{}^a{}_\nu=\partial_\mu e^a_\nu.
\end{equation*}  

The torsion tensor $T^a_{\mu\nu}$ is given by the antisymmetric part of the Weitzenb\"ock connection
\begin{equation} \label{equation-2.2}
	T^a{}_{\mu\nu}=W_\mu{}^a{}_\nu-W_\nu{}^a{}_\mu=\partial_\mu e^a_\nu-\partial_\nu e^a_\mu.
\end{equation}

Two key tensors in TG are the contorsion tensor $K_\mu{}^a{}_\nu$ and the superpotential $S_a{}^{\mu\nu}$. The contorsion tensor $K_\mu{}^\lambda{}_\nu$ is defined as
 \begin{equation} \label{equation-2.3}
	K_\mu{}^\lambda{}_\nu=\frac{1}{2}\left(T^\lambda{}_{\mu\nu}-T_{\mu\nu}{}^\lambda+T_\mu{}^\lambda{}_\nu \right),
\end{equation}
and plays a significant role in establishing the equivalence between TG and GR at the level of field equations. The superpotential $S_\sigma{}^{\mu\nu}$ is given by
\begin{equation} \label{equation-2.4}
	S_\sigma{}^{\mu\nu}=\frac{1}{2}\left(K_\sigma{}^{\mu\nu}-\delta^\mu_\sigma T^\nu+\delta^\nu_\sigma T^\mu \right). 
\end{equation}

The torsion scalar $T$ is constructed through the complete contraction of the torsion tensor with its superpotential
\begin{equation} \label{equation-2.5}
	T=S_\sigma{}^{\mu\nu} T^\sigma{}_{\mu\nu}.
\end{equation}
This scalar serves as the Lagrangian density in the TEGR
\begin{equation} \label{equation-2.6}
	S_\mathrm{TEGR}=-\frac{1}{2\kappa^2}\int d^4 x e T+ \int d^4 x e \mathcal{L}_m,
\end{equation}
where $\kappa^2=8\pi G$ and $\mathcal{L}_m$ is the matter Lagrangian. This quadratic combination encodes the teleparallel equivalent of the Ricci scalar $R$ satisfying the identity
\begin{equation} \label{equation-2.7}
	R=-T+\frac{2}{e}\partial_\mu (eT^\mu)=-T+B,
\end{equation}
where $e=\mathrm{det}\left( e^a_\mu \right) \sqrt{-g}$, and the boundary term is defined as $B=\frac{2}{e}\partial_\mu (eT^\mu)=2\nabla_\mu T^\mu$. 

A natural generalization of the TEGR action is obtained by promoting the torsion scalar $T$ to an arbitrary function $f(T)$, leading to
\begin{equation} \label{equation-2.8}
	S_{f(T)}=-\frac{1}{2\kappa^2}\int d^4 x e f(T)+ \int d^4 x e \mathcal{L}_m.
\end{equation}

Unlike the $f(R)$ framework, which leads to fourth-order field equations, $f(T)$ gravity retains second-order equations of motion. This distinction arises from the relaxed constraints of Lovelock's theorem in teleparallel geometry, where a torsion-based description allows for modifications that avoid ghost instabilities \cite{Krssak 2019}. Nevertheless, the full structure of the theory involves two fundamental geometric scalars: the torsion scalar $T$ and the boundary term $B$. Their relation to the Ricci scalar via the identity $R=-T+B$ motivates the generalization to $f(T, B)$ gravity, which not only encompasses $f(R)$ gravity as a special case but also provides a minimal extension incorporating both second- and fourth-order derivative terms.

In the present work, we consider the $f(T, B)$ action in the form \cite{Caruana 2020}
\begin{equation}\label{equation-2.9}
	S_{f(T, B)}=\frac{1}{2\kappa^2} \int d^4xe(-T + f(T, B))+ \int d^4x e \mathcal{L}_m.
\end{equation}
Note that the TEGR action is recovered when $f(T,B)=0$. Varying this action with respect to the tetrad yields the field equations \cite{Bahamonde 2015,Farrugia 2018}
\begin{equation}\label{equation-2.10}
	\begin{aligned}
	e^\lambda_a \square f_B - e^\sigma_a \nabla^\lambda \nabla_\sigma f_B + \frac{1}{2}B f_B e^\lambda_a+2S^{\mu \lambda}_a (\partial_\mu f_T + \partial_\mu f_B)+\frac{2}{e}(f_T-1)\partial_\mu\left(eS^{\mu \lambda}_a \right)\\
	-2(f_T-1)T^\sigma_{\mu a}S^{\lambda \mu}_\sigma - \frac{1}{2}(-T+f)e^\lambda_a=\kappa^2 \Theta^\lambda_a,
	\end{aligned}
\end{equation}
where $f_T = \partial f /\partial T$, $f_B = \partial f /\partial B$, and $ \Theta^{\lambda}_\nu=e^a_\nu \Theta^\lambda_a $ denotes the standard energy-momentum tensor for matter. These field equations are derived under the assumption of a vanishing spin connection, which is a consistent choice in the context of a flat FLRW cosmology \cite{Bahamonde 2015,Bahamonde 2017,Farrugia 2018}.

The choice of tetrad is 
\begin{equation*}
	e^a_\mu=\mathrm{diag}(1, a(t), a(t), a(t)),
\end{equation*}
where $a(t)$ is the scale factor. This tetrad yields the flat FLRW metric
\begin{equation*}
	ds^2=-dt^2+a(t)(dx^2+dy^2+dz^2),
\end{equation*}
from which the torsion scalar $T$ and the boundary term $B$ are obtained as
\begin{equation} \label{equation-2.11}
	\begin{aligned}
		T=6H^2,\ 
		B=6(3H^2+\dot{H}),
	\end{aligned}
\end{equation}
where the overdot represents a derivative with respect to cosmic time $t$. The corresponding Ricci scalar $R$ for this metric is thus recovered as $R=-T+B=6(\dot{H}+2H^2)$.

With the FLRW metric and the chosen tetrad, the field equations reduce to the modified Friedmann equations
\begin{align}
		3H^2 &=\kappa^2(\rho_m+\rho_{\mathrm{gd}}), \label{equation-2.12} \\ 
		3H^2+2\dot{H}&=-\kappa^2(p_m+p_{\mathrm{gd}}), \label{equation-2.13}
\end{align}
where $\rho_m$ and $p_m$ respectively represent the energy density and pressure of the matter (baryons and dark matter) whose equation of state $\omega_m$ is defined as $p_m=\omega_m \rho_m$, while the geometric dark energy density $\rho_{\mathrm{gd}}$ and the corresponding pressure $p_{\mathrm{gd}}$ are given by 
\begin{align}
		\kappa^2 \rho_{\mathrm{gd}}&=-\frac{1}{2}f + Tf_T + \frac{1}{2}Bf_B-3H\dot{f}_B, \label{equation-2.14} \\
		\kappa^2 p_{\mathrm{gd}}&=\frac{1}{2}f-Tf_T-2\dot{H}f_T-2H\dot{f}_T-\frac{1}{2}Bf_B+\ddot{f}_B, \label{equation-2.15}
\end{align}
The conservation equation of the matter is 
\begin{equation} \label{equation-2.16}
	\dot{\rho}_m+3H\rho_m=0,
\end{equation}
where we have assumed $p_m=0$, then the geometric dark energy also observes the conservation equation \cite{Bahamonde 2017}
\begin{equation} \label{equation-2.17}
	\dot{\rho}_\mathrm{gd}+3H(\rho_{\mathrm{gd}}+p_\mathrm{gd})=0.
\end{equation}
The equation of state for the geometric dark energy $\omega_\mathrm{gd}$ is defined as
\begin{equation} \label{equation-2.18}
	\begin{aligned}
	\omega_\mathrm{gd}=\frac{p_\mathrm{gd}}{\rho_\mathrm{gd}}=-1+\frac{-4\dot{H}f_T-4H\dot{f}_T-6H\dot{f}_B+2\ddot{f}_B}{-f + 2Tf_T + Bf_B-6H\dot{f}_B},
	\end{aligned}
\end{equation}
while the total equation of state is given by
\begin{equation} \label{equation-2.19}
	\omega_\mathrm{tot}=\frac{p_m+p_\mathrm{eff}}{\rho_m+\rho_{\mathrm{eff}}}=-1-\frac{2\dot{H}}{3H^2}.
\end{equation}
The deceleration parameter $q$ is related to $\omega_\mathrm{tot}$ via
\begin{equation} \label{equation-2.20}
	q=-\frac{\ddot{a}}{aH^2}=\frac{1}{2}(1+3\omega_\mathrm{tot}),
\end{equation}
implying that the universe undergoes accelerated expansion when $q<0$, or equivalently, when $\omega_\mathrm{tot}<-\frac{1}{3}$.

In metric teleparallel gravity, with the affine connection $\Gamma^\alpha_{\mu \nu}$ vanishing in FLRW geometry, the boundary term satisfies $C = B$. Under this condition, the modified Friedmann equations of the present $f(T, B)$ model become equivalent to those of the $f(Q, C)$ theory \cite{De 2024, Usman 2024}. This equivalence indicates that the two models may share fundamental features in their cosmological dynamics. Moreover, the dynamical system analysis developed in the following sections for $f(T, B)$ gravity can be directly applied to the $f(Q, C)$ formulation when the affine connection is set to zero.

\section{Dynamical system structure of $f(T, B)$ cosmology} \label{section-3}
In this section, we perform a qualitative analysis of the cosmological dynamics in $f(T, B)$ gravity using the dynamical systems approach. By introducing suitable dimensionless variables, the modified field equations can be reformulated as an autonomous dynamical system. Such a system is characterized by two fundamental components: a state space comprising all possible configurations, and a set of differential equations governing the evolution of trajectories within this space.

The fixed points of the system, defined by the condition $\dot{X} = 0$ for a system of the form $\dot{X} = f(X)$ with $X = (x_1, x_2, \dots, x_n)$, correspond to equilibrium solutions in the cosmological context. These critical points represent distinct cosmological epochs within the $f(T, B)$ framework. Their stability, determined via linear perturbation analysis, dictates the global evolutionary behavior of the universe: stable points act as cosmological attractors characterizing late-time asymptotic states, unstable points correspond to transient phases, and saddle points represent metastable regimes that temporarily influence the dynamics before the system evolves toward an attractor. Through this approach, key epochs in cosmic history -- such as radiation domination, matter domination, and late-time acceleration -- naturally arise as specific critical points in the phase space, revealing how different eras of universe evolution are embedded within the structure of $f(T, B)$ gravity.

To construct the dynamical system for the cosmological model, we introduce the following dimensionless variables:
\begin{equation} \label{equation-3.1}
	\begin{aligned}
	\Omega_m &= \frac{\kappa^2\rho_m}{3H^2}, & \quad \Omega_{\mathrm{gd}} &= \frac{\rho_{\mathrm{gd}}}{6H^2}, & \quad x &= \frac{f}{6H^2}, & \quad u &= f_T, \\
	v &= f_B, & \quad y &= \frac{B}{6H^2}, & \quad z &= \frac{\dot{f}_B}{H}, & \quad \sigma &= yv = \frac{Bf_B}{6H^2}.
	\end{aligned}
\end{equation}

Here, $\Omega_m$ quantifies the relative density of matter in the total effective cosmic fluid, while $\Omega_{\mathrm{gd}} = -x + 2u + \sigma - z$ represents the contribution from geometric dark energy within the $f(T,B)$ framework. The quantities $x$, $u$, $\sigma$, and $z$ characterize different aspects of the geometric dark energy sector. Their relative dominance in the phase space can signal transitions between distinct dark energy-dominated regimes. At fixed points of the dynamical system, the values of these variables help identify the nature of the corresponding cosmological epoch, clarifying the physical role of each component in the evolution governed by the critical points.

Using the dimensionless variables defined in Eq.~\eqref{equation-3.1}, the Friedmann equation~\eqref{equation-2.12} takes the form
\begin{equation} \label{equation-3.2}
	\Omega_m - x + 2u + \sigma - z = 1,
\end{equation}
while Eq.~\eqref{equation-2.13} becomes
\begin{equation} \label{equation-3.3}
	\frac{\ddot{f}_B}{H^2} = -3\Omega_m + 2(y - 3)(u - 1) + 3z + 2\frac{\dot{f}_T}{H}.
\end{equation}
The term $\dot{f}_T/H$ can be expanded as
\begin{equation} \label{equation-3.4}
	\frac{\dot{f}_T}{H} = 2(y - 3)Tf_{TT} + \frac{\dot{B}}{6H^2}Tf_{TB},
\end{equation}
where $f_{TT} = \partial^2 f / \partial T^2$ and $f_{TB} = \partial^2 f / (\partial T \partial B)$.

Furthermore, from the dynamical variables $y$ and $z$, we derive the following relations:
\begin{align}
	\frac{\dot{H}}{H^2} &= y - 3, \label{equation-3.5} \\
	\frac{\dot{B}}{6H^3} &= \frac{z - 2(y - 3)Tf_{BT}}{Tf_{BB}}, \label{equation-3.6}
\end{align}
in which $f_{BT} = \partial^2 f / (\partial B \partial T)$, $f_{BB} = \partial^2 f / \partial B^2$, and it is assumed that $f_{BB} \neq 0$.

Finally, the dark energy equation of state $\omega_{\mathrm{gd}}$, the total equation of state $\omega_{\mathrm{tot}}$, and the deceleration parameter $q$ are given by
\begin{equation} \label{equation-3.7}
	\omega_{\mathrm{gd}} = \frac{2 - 3y}{3(1 - \Omega_m)}, \quad
	\omega_{\mathrm{tot}} = 1 - 2/3y, \quad
	q = 2 - y.
\end{equation}

To construct the cosmological dynamical system, we introduce the independent variable $N \equiv \ln a$, commonly used in expanding cosmological scenarios ($H > 0$) but inapplicable in bouncing models where $H = 0$ at the bounce epoch \cite{Caruana 2020}. The field equations of $f(T, B)$ gravity can then be expressed as the following autonomous dynamical system:
\begin{equation} \label{equation-3.8}
	\begin{aligned}
		\Omega_m'&=\Omega_m(3-2y),\\
		x'&=2(y-3)(u-x)+v\frac{\dot{B}}{6H^3},\\
		u'&=\frac{\dot{f}_T}{H},\\
		\sigma'&=yz-2\sigma(y-3)+v\frac{\dot{B}}{6H^3},\\
		y'&=-2y(y-3)+\frac{\dot{B}}{6H^3},\\
		v'&=z,\\
		z'&=-(y-3)z+\frac{\ddot{f}_B}{H^2},
	\end{aligned}
\end{equation}
where the prime symbol $ ' $ denotes differentiation with respect to $N = \ln a$. Using the variables defined in Eq.~\eqref{equation-3.1} and the relations in Eqs.~\eqref{equation-3.2} and~\eqref{equation-3.3}, the system can be reduced to:
\begin{equation} \label{equation-3.9}
	\begin{aligned}
		\Omega_m'&=\Omega_m(3-2y),\\
		x'&=2(y-3)(u-x)+v\frac{\dot{B}}{6H^3},\\
		u'&=\frac{\dot{f}_T}{H},\\
		y'&=-2y(y-3)+\frac{\dot{B}}{6H^3},\\
		v'&=-1+\Omega_m-x+2u+yv.
	\end{aligned}
\end{equation}

Upon specifying the functional form of $f(T, B)$, the dynamical system presented above becomes fully autonomous, in contrast to approaches that rely on the parameterization $\lambda = \ddot{H}/H^3$ as used in \cite{Franco 2020, Samaddar 2023, Kadam 2023, Odintsov 2017}. In the subsequent sections, we focus on two specific $f(T, B)$ models introduced in \cite{Bahamonde 2017}: the power-law model
$$f(T, B) = c_1 T^\alpha B^\beta,$$
and the mixed power-law model
$$f(T, B) = c_2 T^\alpha + c_3 B^\beta,$$
where $c_1, c_2, c_3, \alpha, \beta$ are constant parameters of the cosmological model. 
Both of these representative and well-motivated prototype functional forms are chosen because they can naturally reduce to GR or to other established modified gravity theories, such as $f(T) $ or $f(R) $ gravity, within specific parameter limits, thereby ensuring theoretical consistency. They also serve complementary aims: the multiplicative power-law form is intended to explore novel dynamical effects that stem from a non-trivial coupling between the torsion scalar $T $ and the boundary term $B $, an interaction inherently absent in additive or pure $f(R) $ models. In contrast, the additive power-law form enables a clear separation and comparative assessment of the individual contributions of $T $ and $B $ to cosmic evolution. Furthermore, the power-law ansatz yields homogeneous terms in the resulting Friedmann equations, which significantly facilitates the search for exact scaling solutions and the construction of a closed autonomous dynamical system.
It should also be noted that other forms of $f(T,B) $ may be considered in future studies, such as $f(T,B)=A_0 + A_1 T + A_2 T^2 + A_3 B + A_4 T B $ \cite{Franco 2020,Farrugia 2018}, $f(T,B) = \xi T+ \alpha B\ln B $ \cite{Kadam 2023}, as well as $f(T,B) = B g(T) $ and $f(T,B) = T g(B) $ \cite{Caruana 2020}. By systematically analyzing the two foundational ansätze selected here, we aim to map the key dynamical features of $f(T,B)$ cosmology and establish a benchmark for future studies involving more complex functional dependencies.

\section{Cosmological dynamics of two $f(T, B)$ models} \label{section-4}
\subsection{Power law model $f(T, B)=c_1 T^\alpha B^\beta$} \label{section-4.1}
We first consider the power law model $f(T, B)=c_1 T^\alpha B^\beta$. For Eq.~\eqref{equation-3.6} to be well-defined, the condition $f_{BB} \neq 0$ must be satisfied, which requires $\beta \neq 0, 1$ and $c_1 \neq 0$. In this case, the dynamical variables $u$ and $v$ can be written as
\begin{equation} \label{equation-4.1}
	u = \alpha x, \quad v = \beta\frac{x}{y}.
\end{equation}

Substituting these into Eq.~\eqref{equation-3.6} yields
\begin{equation} \label{equation-4.2}
	\frac{\dot{B}}{6H^3}=\frac{y^2(-1+\Omega_m+(-1+2\alpha+\beta))-2\alpha \beta xy(y-3)}{(\beta-1)\beta x}.
\end{equation}

The cosmological dynamical system for the power law model then reduces to the autonomous form
\begin{equation} \label{equation-4.3}
	\begin{aligned}
		\Omega_m'&=\Omega_m(3-2y),\\
		x'&=2(\alpha-1)(y-3)x+\frac{y(-1+\Omega_m+(-1+2\alpha+\beta))-2\alpha \beta x(y-3)}{\beta-1},\\
		y'&=-2y(y-3)+\frac{y^2(-1+\Omega_m+(-1+2\alpha+\beta))-2\alpha \beta xy(y-3)}{(\beta-1)\beta x}.
	\end{aligned}
\end{equation}

The autonomous system~\eqref{equation-4.3} admits a unique fixed point, denoted as $P^1_{\mathrm{gd}}$. Its coordinates and the corresponding cosmological parameters are summarized in Table~\ref{table-1}. The eigenvalues of the Jacobian matrix evaluated at this point, denoted as $\{e_{a1}, e_{a2}, e_{a3}\}$, are given by
\begin{equation*}
	\left\{-3, -\frac{3}{2} - \frac{3 \sqrt{A_1}}{2\beta(\beta-1)(2\alpha+\beta-1)}, -\frac{3}{2} + \frac{3 \sqrt{A_1}}{2\beta(\beta-1)(2\alpha+\beta-1)} \right\},
\end{equation*}
where
\begin{equation*}
	\begin{aligned}
	A_1=& \beta (\beta -1)(2\alpha +\beta -1)^2(8+16\alpha^2+24\alpha(\beta -1)-17\beta +9\beta^2).
	\end{aligned}
\end{equation*}

\begin{figure}[htbp] 
	\centering
	\subfigure[]{\includegraphics[scale=0.46]{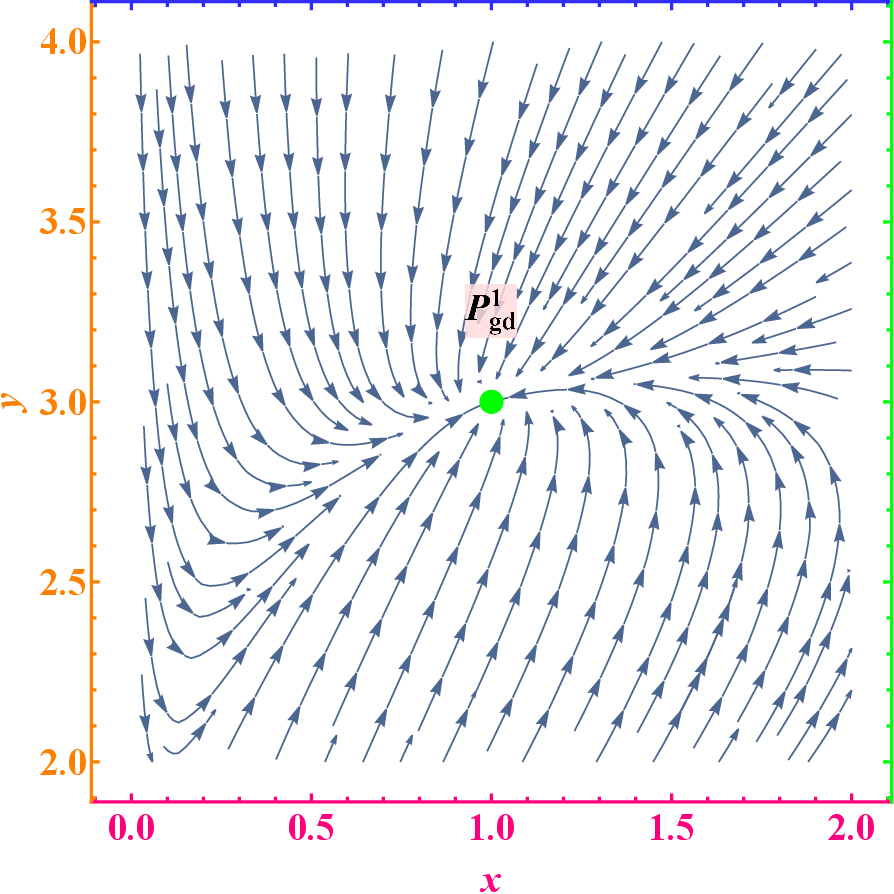}
		\label{figure-2a}}
	\subfigure[]{\includegraphics[scale=0.46]{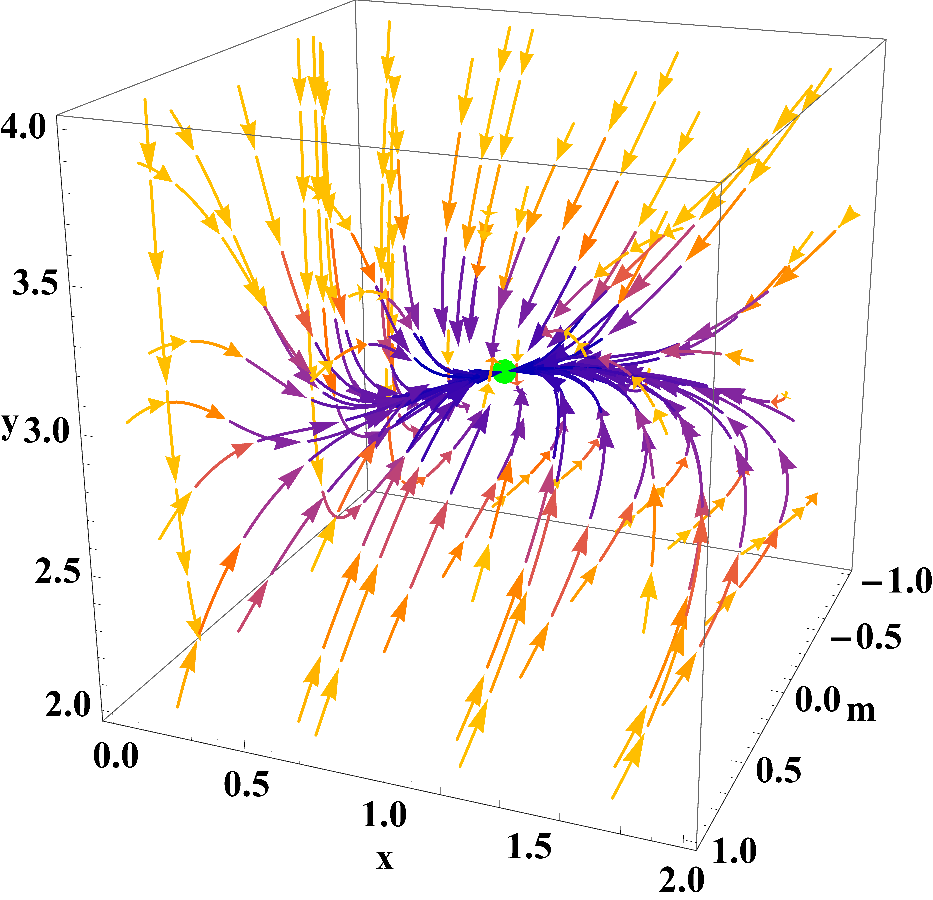}
		\label{figure-2b}}
	\caption{Phase space flow of the model $f(T, B)=c_1 T^\alpha B^\beta$ for $(\alpha_1, \beta_1)=(3,-4)$.}
	\label{figure-1}
\end{figure}

\begin{figure}[htbp] 
	\centering
	\includegraphics[scale=0.46]{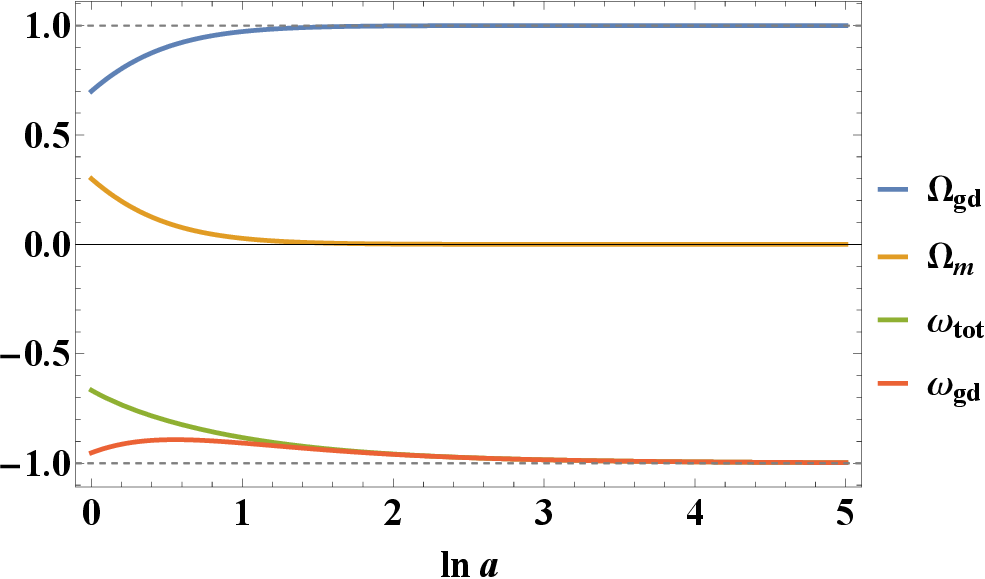}
	\caption{Evolution of cosmological parameters for $(\alpha_1, \beta_1)=(3,-4)$ and initial conditions $(\Omega_{m_0}, x_0, y_0)=(0.3, 0.3, 2.5)$ .}
	\label{figure-2}
\end{figure}

Table~\ref{table-2} summarizes the existence and linear stability conditions for $P^1_{\mathrm{gd}}$, along with its acceleration behavior, where the symbols $\wedge$ and $\vee$ denote logical ``and'' and ``or'', respectively. For the specific parameter values $\alpha = 3$ and $\beta = -4$, the fixed point $P^1_{\mathrm{gd}}$ is a stable node. The phase space stream plot of the model for this parameter set is shown in Figure~\ref{figure-1}, while the evolution of the corresponding cosmological parameters is displayed in Figure~\ref{figure-2}.

The fixed point $P^1_{\mathrm{gd}}$ corresponds to a cosmological epoch dominated by geometric dark energy, characterized by a de Sitter expansion $a \sim e^{H_0 t}$ and a total equation of state $\omega_{\mathrm{tot}} = -1$. At this point, the geometric dark energy components $x$, $u$, and $\sigma$ collectively sustain the accelerated expansion. Owing to its stable nodal behavior for suitable parameter choices $(\alpha, \beta)$, this fixed point provides a viable mechanism for explaining the late-time cosmic acceleration within the power-law $f(T, B)$ framework.
\begin{table}[htbp]
	\centering
	\caption{Fixed point of dynamical system (\ref{equation-4.3})}
		\begin{tabular}{c c c c c c c c c c c c}
			\toprule[1pt]
			\textbf{Point} &  $\Omega_m$ & $ \Omega_\mathrm{gd} $ &  $x$ &  $y$ & $u$ & $v$ & $\sigma$ & $z$ & $\omega_\mathrm{tot}$ & $H$ & $a$  \\
			\midrule[1pt]
			$P^1_{\mathrm{gd}}$ &  $0$ & $1$ & $\frac{1}{ 2\alpha+\beta-1 }$ & $3$ & $\frac{\alpha}{ 2\alpha+\beta-1 }$ & $\frac{\beta}{ 3(2\alpha+\beta-1)}$& $\frac{\beta}{2\alpha+\beta-1}$ & $0$ & $-1$ & $H_0$ & $e^{H_0t}$ \\
			\bottomrule[1pt]
	\end{tabular}
	\label{table-1}
\end{table}
\begin{table}[htbp]
	\centering
	\caption{Summary of existence, stability, and acceleration properties for the fixed point $P^1_{\mathrm{gd}}$}
	\resizebox{\linewidth}{!}{
	\begin{tabular}{c c l c }
		\toprule[1pt]
		\textbf{Point} & \textbf{Existence} & \textbf{Stability} & \textbf{Acceleration}  \\
		\midrule[1pt]
		$P^1_{\mathrm{gd}}$ & $2\alpha +\beta \neq 1 $ &  \makecell[l]{ stable for ($\alpha\leq 0 \wedge (0 \leq \beta \leq 1 \vee 1-\alpha< \beta \leq 1-2\alpha ) )$ \\ $\vee (0<\alpha<\frac{1}{2} \wedge (0\leq \beta \leq 1-2\alpha \vee 1-\alpha <\beta\leq 1) )$  \\  $\vee  (\alpha =\frac{1}{2} \wedge (\beta=0 \vee \frac{1}{2}<\beta\leq 1 )$ \\  $ \vee (\frac{1}{2}<\alpha<1 \wedge (1-2\alpha \leq \beta <0 \vee 1-\alpha < \beta \leq 1 ) ) )$ \\  $ \vee ( \alpha =1 \wedge -1 \leq \beta \leq 1 ) $  \\   $ \vee (\alpha >1 \wedge ( 1-2\alpha \leq \beta <1-\alpha \vee 0\leq \beta \leq 1 )) $ } & always \\
		\bottomrule[1pt]
	\end{tabular}}
	\label{table-2}
\end{table}

\subsection{Mixed power law model $f(T, B)=c_2 T^\alpha + c_3 B^\beta$} \label{section-4.2}
We now consider the mixed power law model $f(T, B) = c_2 T^\alpha + c_3 B^\beta$. Here, the condition $f_{BB} \neq 0$ is satisfied provided that $\beta \neq 0, 1$ and $c_3 \neq 0$. In this model, the variable $x$ and the term $\dot{B}/(6H^3)$ can be written as
\begin{align}
	x &= \frac{1}{\alpha}u + \frac{1}{\beta}yv, \label{equation-5.1} \\
	\frac{\dot{B}}{6H^3} &= \frac{y\left(-1 + \Omega_m + 2u + yv - \frac{1}{\alpha}u - \frac{1}{\beta}yv\right)}{(\beta - 1)v}. \label{equation-5.2}
\end{align}
The corresponding autonomous dynamical system takes the form
\begin{equation} \label{equation-5.3}
	\begin{aligned}
		\Omega_m' &= \Omega_m(3 - 2y), \\
		u' &= 2(\alpha - 1)(y - 3)u, \\
		y' &= -2(y - 3)y + \frac{y\left(-1 + \Omega_m + 2u + yv - \frac{1}{\alpha}u - \frac{1}{\beta}yv\right)}{(\beta - 1)v}, \\
		v' &= -1 + \Omega_m + 2u + yv - \frac{1}{\alpha}u - \frac{1}{\beta}yv.
	\end{aligned}
\end{equation}
This system possesses two fixed points, denoted as $P^i_{\mathrm{gd}}$ with coordinates $(\Omega^i_m, u^i, y^i, v^i)$ for $i = 2, 3$.

The fixed point $P^2_{\mathrm{gd}}$ is characterized by the coordinates
\begin{equation*}
	P^2_{\mathrm{gd}} = \left( 0,\; \frac{\alpha(\beta + 3(1 - \beta)v_*)}{(2\alpha - 1)\beta},\; 3,\; v_* \right),
\end{equation*}
where $v_* \in \mathbb{R}$ and $v_* \neq 0$. More precisely, $P^2_{\mathrm{gd}}$ corresponds to a line of equilibrium points. The eigenvalues of the linearized system at this point are given by
\begin{equation*}
	\left\{0,\; -3,\; -\frac{3}{2} - \frac{\sqrt{3 A_2}}{2 (2\alpha - 1)(\beta - 1)\alpha \beta v_*},\; -\frac{3}{2} + \frac{\sqrt{3 A_2}}{2 (2\alpha - 1)(\beta - 1)\alpha \beta v_*} \right\},
\end{equation*}
with
\begin{equation*}
	A_2 = v_* \alpha^2 \beta (2\alpha - 1)^2 (\beta - 1) \left[8\beta (\alpha - 1) - 3v_*(8\alpha - 9\beta)(\beta - 1)\right].
\end{equation*}
At this fixed point, the cosmological parameters satisfy $\Omega_m = 0$ and $\Omega_{\mathrm{gd}} = 1$, indicating a universe dominated by geometric dark energy. As a de Sitter point with a scale factor evolving as $a \sim e^{H_0 t}$, $P^2_{\mathrm{gd}}$ provides a potential explanation for the current cosmic acceleration within the model, provided it acts as a stable attractor.

The fixed point $P^3_{\mathrm{gd}}$ is located at
\begin{equation*}
	P^3_{\mathrm{gd}} = \left(0,\ 0,\ 3,\ \frac{\beta}{3(\beta - 1)} \right).
\end{equation*}
This point belongs to the equilibrium line $P^2_{\mathrm{gd}}$, corresponding to the specific case where $v_* = \frac{\beta}{3(\beta - 1)}$. The eigenvalues of the linearization at $P^3_{\mathrm{gd}}$ are
\begin{equation*}
	\left\{0,\ -3,\ -\frac{3}{2} + \frac{3\sqrt{9\alpha^2\beta^4 - 8\alpha^2\beta^3}}{2\alpha\beta^2},\ -\frac{3}{2} - \frac{3\sqrt{9\alpha^2\beta^4 - 8\alpha^2\beta^3}}{2\alpha\beta^2} \right\}.
\end{equation*}
Like $P^2_{\mathrm{gd}}$, this point also corresponds to a phase of geometric dark energy dominance and exponential expansion of the universe. If linearly stable, it could provide a mechanism for late-time cosmic acceleration.

\begin{figure}[htbp] 
	\centering
	\subfigure[]{\includegraphics[scale=0.46]{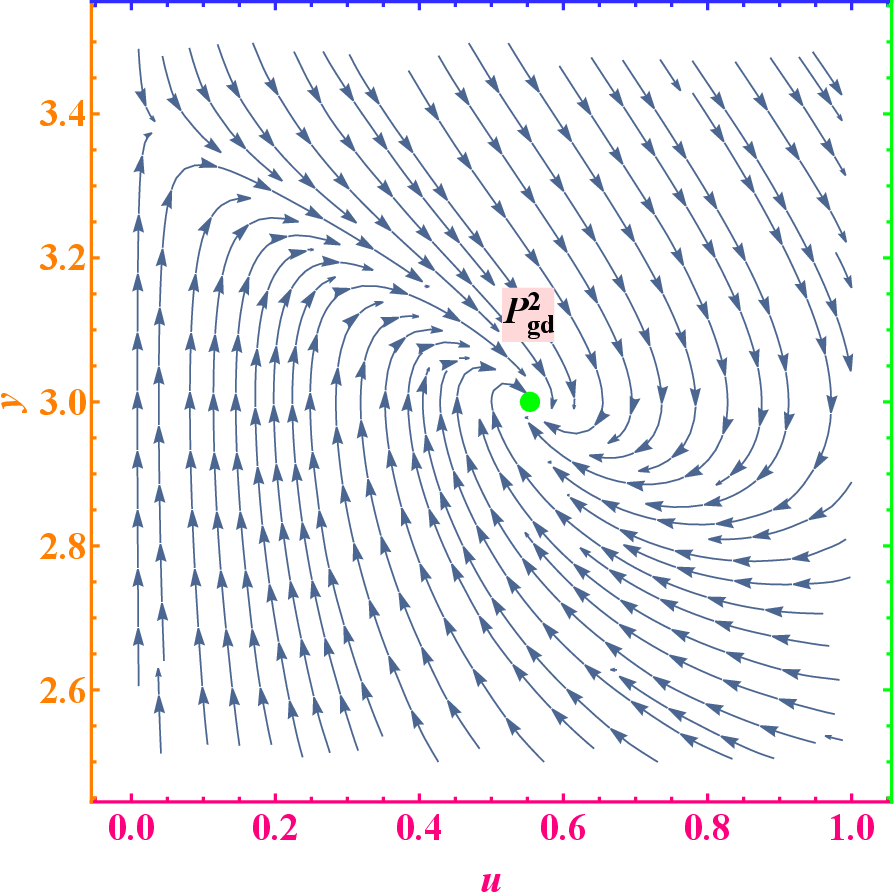}
		\label{figure-3a}}
	\subfigure[]{\includegraphics[scale=0.46]{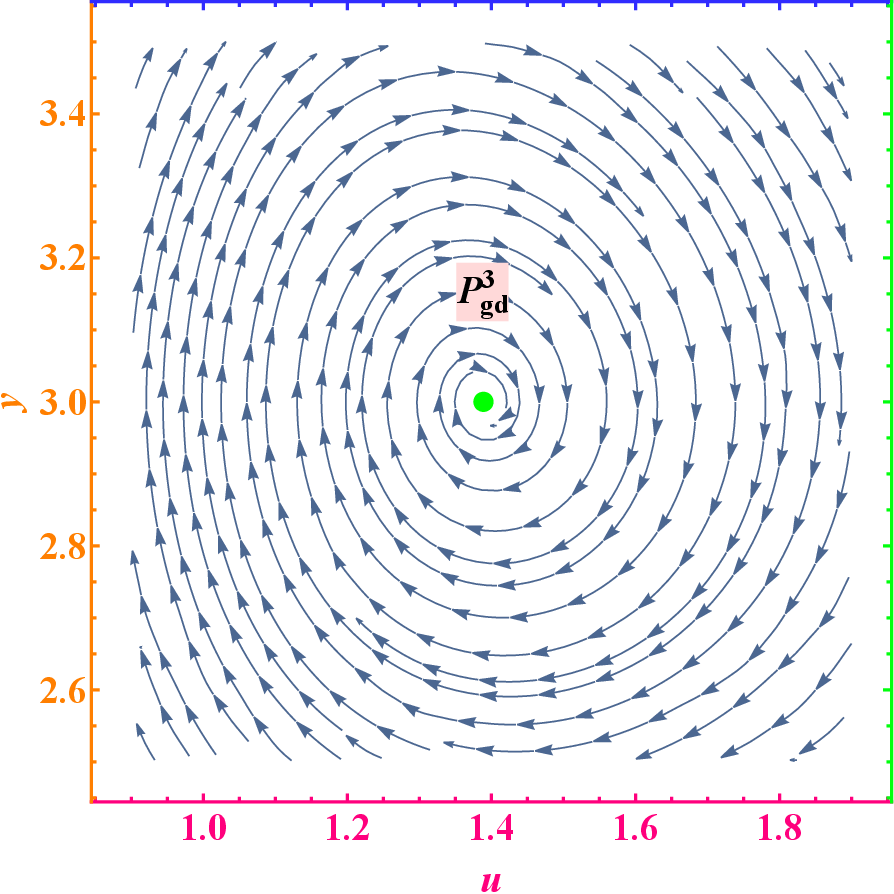}
		\label{figure-3b}}
	\subfigure[]{\includegraphics[scale=0.46]{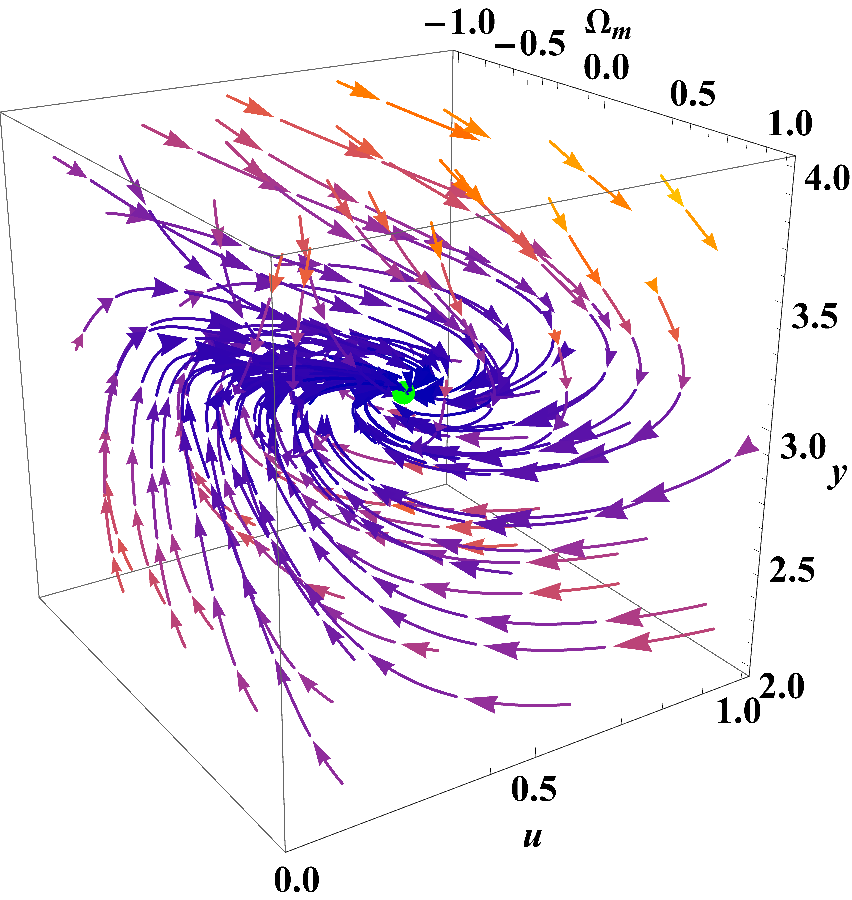}
		\label{figure-3c}}
	\subfigure[]{\includegraphics[scale=0.46]{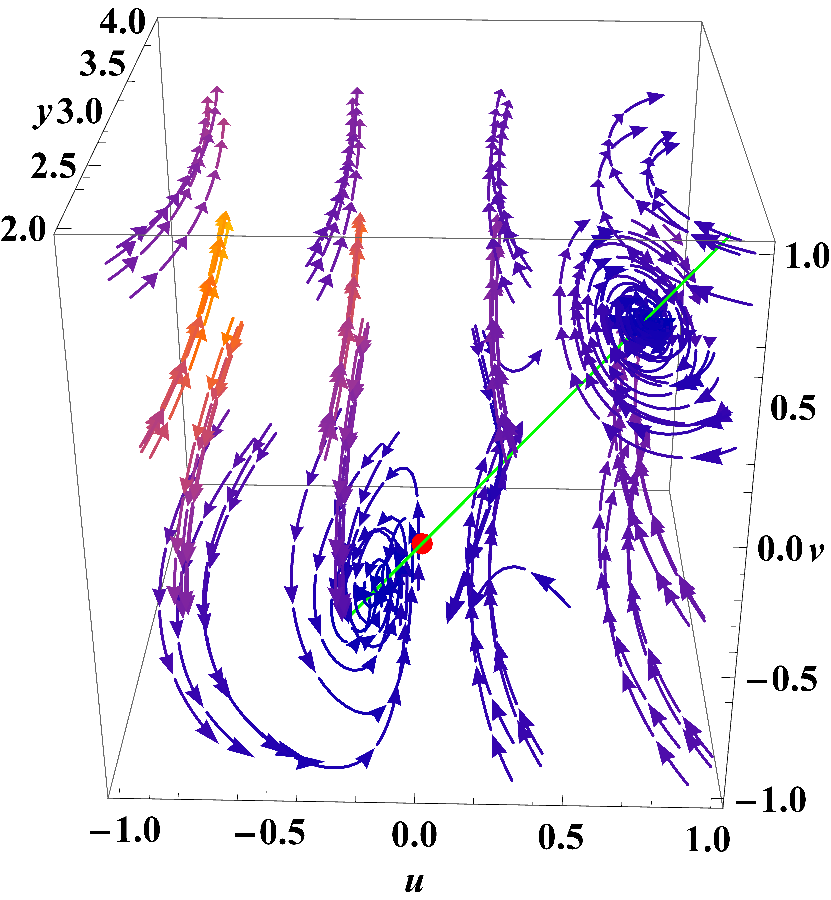}
		\label{figure-3d}}
	\caption{Phase space flow of the mixed power-law model under different parameters: (a), (c): $(\alpha_2, \beta_2, v_2)=(5,-1000, 0.00125)$; (b): $(\alpha_3, \beta_3, v_3)=(5, 2/3, 1)$; (d): $(\alpha_3, \beta_3)=(5, 2/3)$.}	
	\label{figure-3}
\end{figure}

The coordinates and cosmological parameters of both fixed points are summarized in Table~\ref{table-3}, while their existence, stability, and acceleration properties are listed in Table~\ref{table-4}, where $\tilde{v} = \beta(\alpha - 1)/(3(\beta - 1)(\alpha - \beta))$. In the mixed power-law model, both obtained fixed points exhibit one zero eigenvalue and are therefore non-hyperbolic. The central manifold theorem was attempted to assess their stability; however, after decomposing the system into linear and nonlinear components, it was found that the nonlinear terms do not vanish in the vicinity of the equilibrium, thus precluding definitive stability conclusions through this method. Consequently, we specify the condition that the real parts of the eigenvalues $e_{j3}$, $e_{j4}$ (with $j = b, c$) must be negative in Table~\ref{table-4}, and present the stability behavior of both points via phase portrait analysis in Figure~\ref{figure-3}.
\begin{table}[htbp]
	\centering
	\caption{Fixed points of dynamical system (\ref{equation-5.3})}
	\begin{tabular}{c c c c c c c c c c c c}
		\toprule[1pt]
		\textbf{Point} & $\Omega_m$ & $ \Omega_\mathrm{gd} $ & $x$ & $y$ & $u$ & $v$ & $\sigma$ & $z$ & $\omega_\mathrm{tot}$ & $H$ & $a$  \\
		\midrule[1pt]
		$P^2_{\mathrm{gd}}$ &  $0$ & $1$ & $\frac{\beta +3v_*(2\alpha-\beta)}{ (2\alpha -1)\beta }$ & $3$ & $\frac{\alpha \beta -3\alpha v_*(\beta-1)}{ (2\alpha-1)\beta }$ & $v_*$& $3v_*$ & $0$ & $-1$ & $H_0$ & $e^{H_0t}$ \\
		$P^3_{\mathrm{gd}}$ &  $0$ & $1$ & $\frac{1}{\beta-1 }$ & $3$ & $0$ & $\frac{\beta}{ 3(\beta-1)}$& $\frac{\beta}{\beta-1}$ & $0$ & $-1$ & $H_0$ & $e^{H_0t}$ \\
		\bottomrule[1pt]
	\end{tabular}
	\label{table-3}
\end{table}
\begin{table}[htbp]
	\centering
	\caption{Existence, stability, and acceleration properties of $P^2_{\mathrm{gd}}$ and $P^3_{\mathrm{gd}}$}
	\resizebox{\linewidth}{!}{
		\begin{tabular}{c c l c }
			\toprule[1pt]
			\textbf{Point} & \textbf{Existence} & \textbf{Stability} $( e_{j3}, e_{j4}<0, j=b,c  )$ & \textbf{Acceleration}  \\
			\midrule[1pt]
			$P^2_{\mathrm{gd}}$ & \makecell[l]{$\beta \neq 0$ \\ $\wedge \alpha \neq \frac{1}{2}$ \\  $ \wedge v_* \neq 0$ } &  \makecell[l]{ 
				($\alpha < 0 \wedge ((\beta \leq \alpha \wedge \tilde{v}< v_* < 0 ) \vee (\beta=\alpha \wedge v_*<0) \vee ( \alpha<\beta <0 \wedge (v_*<0 \vee v_*> \tilde{v}) )$ \\
				 $ \vee (0<\beta <1 \wedge \tilde{v} <v_*<0) \vee (\beta>1 \wedge 0<v_*< \tilde{v}) ) )$  \\ 
				 $ \vee (0<\alpha <\frac{1}{2} \wedge ( ( \beta <0 \wedge \tilde{v} <v_*<0 ) \vee ( 0<\beta <\alpha \wedge ( v_*<0 \vee v_*> \tilde{v} ) \vee (\beta =\alpha \wedge v_*<0) $ \\  
				 $ \vee ( \alpha<\beta <1 \wedge \tilde{v} <v_*<0 ) \vee ( \beta >1 \wedge 0<v_*< \tilde{v} ) )  )  $  \\  
				 $ \vee (\frac{1}{2}<\alpha <1 \wedge ( (\beta<0 \wedge \tilde{v} <v_*<0 ) \vee ( 0<\beta <\alpha \wedge ( v_*<0 \vee v_*> \tilde{v} ) ) \vee (\beta =\alpha  \wedge v_*<0) $  \\   
				 $ \vee (\alpha <\beta <1 \wedge \tilde{v} <v_*<0 )  \vee ( \beta >1 \wedge 0<v_*< \tilde{v} )  ) ) $   \\  
				 $ \vee (\alpha =1 \wedge 0<\beta <1 \wedge ( v_*<0 \vee v_*>0 ) )$  \\   
				 $ \vee ( \alpha >1 \wedge ( (\beta <0 \wedge  0<v_*< \tilde{v}) \vee (0<\beta <1 \wedge (v_*< \tilde{v} \vee v_*>0 )) $ \\ 
				 $ \vee (1<\beta <\alpha  \wedge (v_*<0 \vee v_* > \tilde{v} ))   \vee  (\beta =\alpha \wedge v_*<0) \vee (\beta >\alpha  \wedge \tilde{v} <v_* <0)  ) )  $ } & always \\
			$P^3_{\mathrm{gd}}$ & $\beta \neq 1 $ &  \makecell[l]{ $\alpha\neq 0 \wedge  0<\beta <1 $ } & always \\
			\bottomrule[1pt]
	\end{tabular}}
	\label{table-4}
\end{table}

\begin{figure}[htbp] 
	\centering
	\subfigure[]{\includegraphics[scale=0.45]{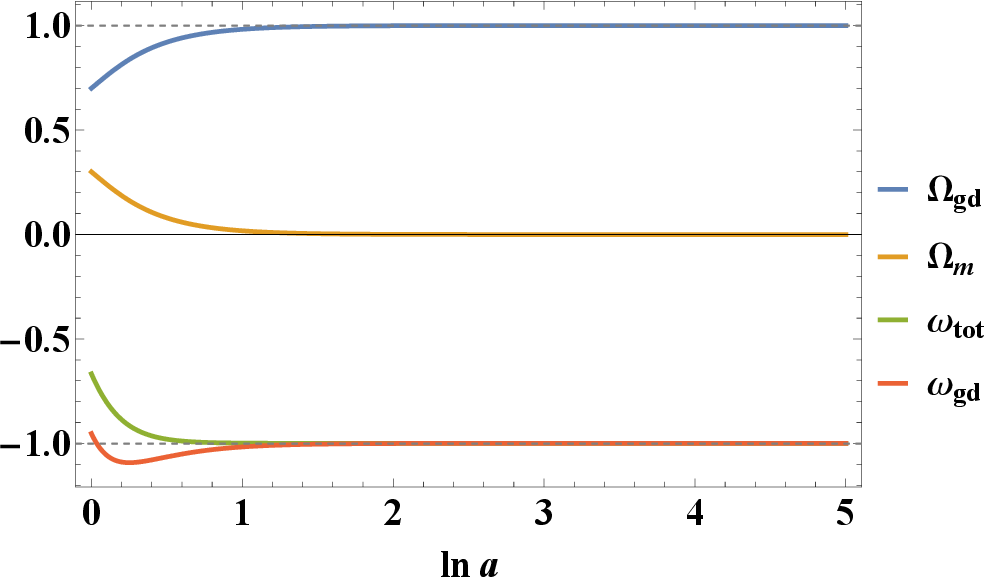}
		\label{figure-4a}}
	\subfigure[]{\includegraphics[scale=0.45]{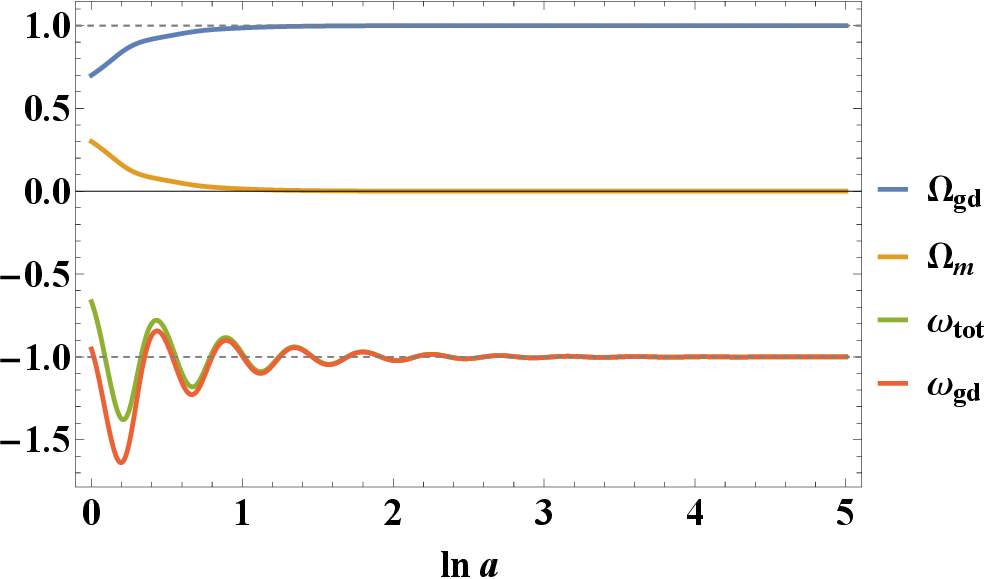}
		\label{figure-4b}}
	\caption{Evolution of cosmological parameters for the mixed power-law model from initial conditions $(\Omega_{m_0}, u_0, y_0, v_0)=(0.3, 1, 2.5, 1)$. (a): $(\alpha_2, \beta_2)=(5,-1000)$; (b): $(\alpha_3, \beta_3)=(5,2/3)$. }
	\label{figure-4}
\end{figure}

We perform numerical analysis for two representative parameter pairs: $(\alpha_2, \beta_2) = (5, -1000)$ and $(\alpha_3, \beta_3) = (5, 2/3)$, as illustrated in Figures~\ref{figure-3} and~\ref{figure-4}. In the $y$-$u$ plane, $P^2_{\mathrm{gd}}$ is stable under both parameter choices and exhibits spiral dynamics for $(5, -1000)$. The cosmological parameters $\Omega_m$, $\Omega_{\mathrm{gd}}$, $\omega_{\mathrm{tot}}$, and $\omega_{\mathrm{gd}}$ all asymptotically approach the same final state $(0, 1, -1, -1)$. However, for the pair $(5, 2/3)$, the evolution of $\omega_{\mathrm{tot}}$ and $\omega_{\mathrm{gd}}$ shows distinct damped oscillations before settling to the de Sitter attractor.

\section{Statefinder diagnostic} \label{section-5}
This section applies the statefinder diagnostic to the two cosmological models introduced above. The statefinder parameters $ r $ and $ s $, first proposed in \cite{Sahni 2003,Alam 2003}, provide a useful tool for distinguishing between different dark energy scenarios. These dimensionless quantities are defined in terms of the scale factor and its higher-order derivatives as \cite{Sahni 2003}:
\begin{equation} \label{equation-6.1}
	r = \frac{\dddot{a}}{aH^3}, \quad 
	s = \frac{r - 1}{3(q - \frac{1}{2})}.
\end{equation}
The statefinder diagnostic offers a practical framework for classifying alternative dark energy models. Even when two models predict similar expansion histories, their evolutionary trajectories in the $r$-$s$ plane can reveal clear differences. Characteristic values of $ \{r, s\} $ for several standard dark energy models include \cite{Sahni 2003,Alam 2003}:
\begin{itemize}
	\item[$\bullet$] $ \{r = 1, s = 1\} $: Standard cold dark matter (SCDM) model
	\item[$\bullet$] $ \{r = 1, s = 0\} $: $\Lambda$CDM model
	\item[$\bullet$] $ \{r > 1, s < 0\} $: Chaplygin gas model
	\item[$\bullet$] $ \{r < 1, s > 0\} $: Quintessence model
\end{itemize}

The application of statefinder diagnostics within a dynamical systems framework proves particularly effective in distinguishing between different acceleration regimes, as demonstrated in recent studies of dark energy models \cite{Panotopoulos 2020,Shukla 2025}. In present work, for the power-law model $f(T, B) = c_1 T^\alpha B^\beta$, the statefinder parameters take the form
\begin{equation} \label{equation-6.2}
	\begin{aligned}
		r&=10-3y+\frac{-2\alpha \beta xy(y-3)+y^2[\Omega_m+(2\alpha +\beta -1)x-1]}{\beta (\beta -1)x}, \\
		s&=\frac{-6\beta (\beta-1)(y-3)x -4\alpha \beta xy(y-3)+2y^2[\Omega_m+(2\alpha +\beta -1)x-1)]}{3\beta (\beta -1)(3-2y)x},
	\end{aligned}
\end{equation}
whereas for the mixed power-law model $f(T, B) = c_2 T^\alpha + c_3 B^\beta$, they are given by
\begin{equation} \label{equation-6.3}
	\begin{aligned}
		r&=10-3y+\frac{y(-1+\Omega_m+2u+yv-\frac{u}{\alpha}-\frac{yv}{\beta})}{(\beta -1)v}, \\
		s&=\frac{-6(\beta -1)(y-3)v+2y(-1+\Omega_m+2u+yv-\frac{u}{\alpha}-\frac{yv}{\beta})}{3(\beta -1)(3-2y)v}.
	\end{aligned}
\end{equation}
In both cases, the deceleration parameter is expressed as $q = 2 - y$.

\begin{figure}[htbp] 
	\centering
	\subfigure[]{\includegraphics[scale=0.49]{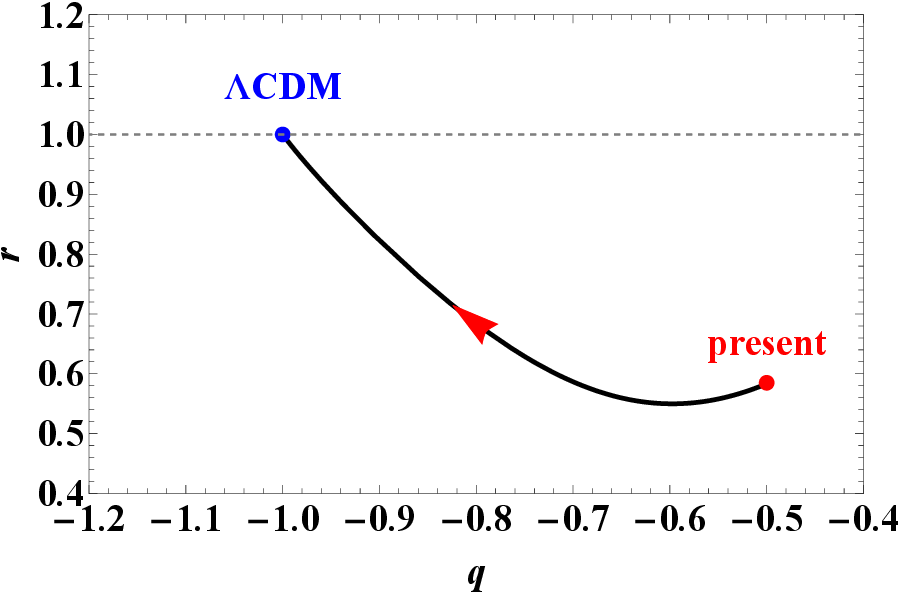}
		\label{figure-5a}}
	\subfigure[]{\includegraphics[scale=0.49]{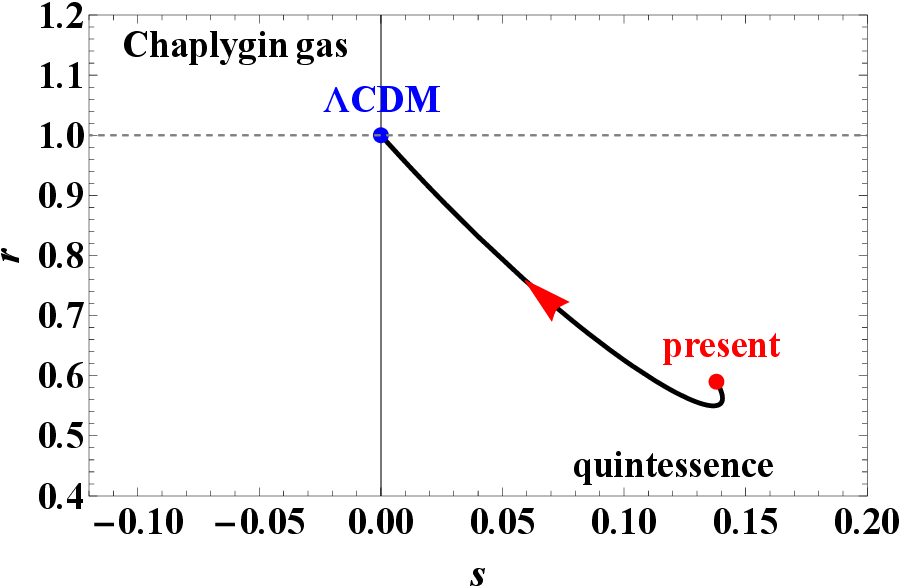}
		\label{figure-5b}}
	\caption{Evolution of statefinder parameters in the $r$-$q$ and $r$-$s$ planes for the model $f(T, B) = c_1 T^\alpha B^\beta$, with initial values $(\Omega_{m_0}, x_0, y_0) = (0.3, 0.3, 2.5)$ and parameter choice $(\alpha_1, \beta_1) = (3, -4)$.}
	\label{figure-5}
\end{figure}

\begin{figure}[htbp] 
	\centering
	\subfigure[]{\includegraphics[scale=0.49]{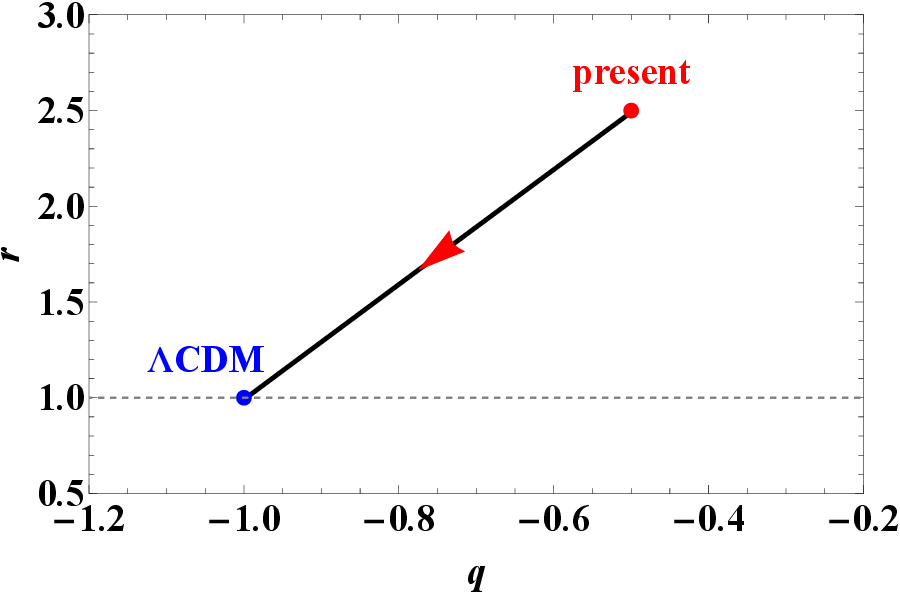}
		\label{figure-6a}}
	\subfigure[]{\includegraphics[scale=0.49]{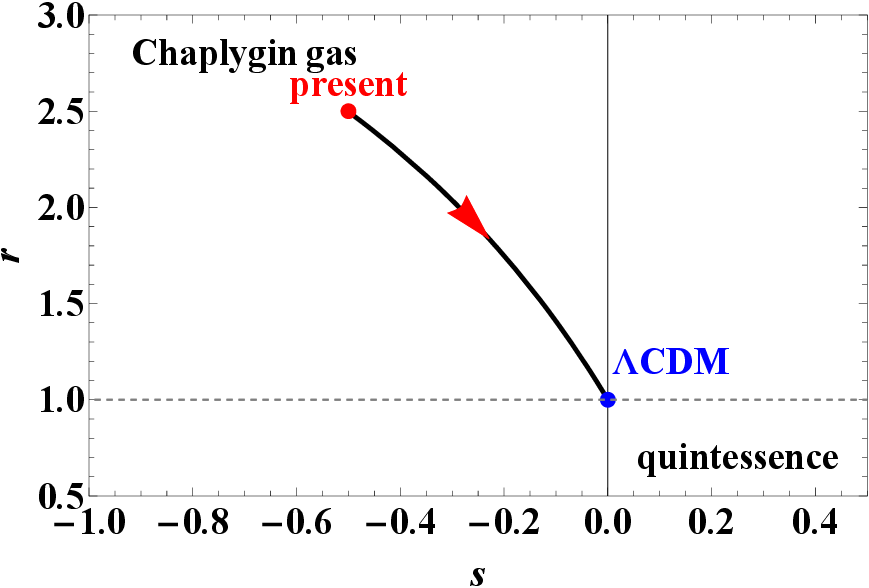}
		\label{figure-6b}}
	\subfigure[]{\includegraphics[scale=0.48]{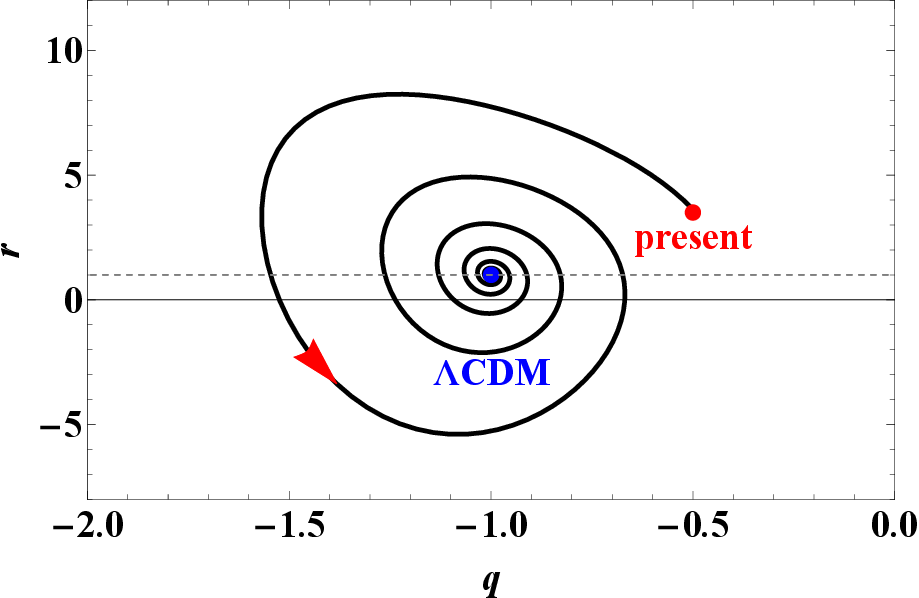}
		\label{figure-6c}}
	\subfigure[]{\includegraphics[scale=0.48]{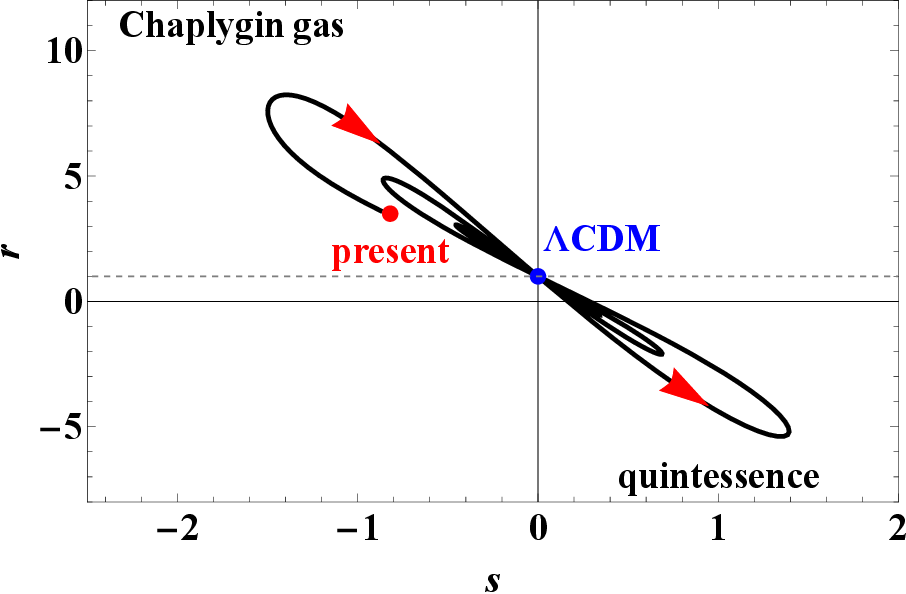}
		\label{figure-6d}}
	\caption{Evolution of statefinder parameters in the $r$-$q$ and $r$-$s$ planes for the model $f(T, B) = c_2 T^\alpha + c_3 B^\beta$, with initial values $(\Omega_{m_0}, u_0, y_0, v_0) = (0.3, 1, 2.5, 1)$. Panels (a) and (b): $(\alpha_2, \beta_2) = (5, -1000)$; panels (c) and (d): $(\alpha_3, \beta_3) = (5, 2/3)$.}
	\label{figure-6}
\end{figure}

The evolutionary trajectories of the statefinder parameters $r$ and $s$ are illustrated in the $r$-$q$ and $r$-$s$ planes: Figure~\ref{figure-5} corresponds to the power-law model $f(T, B) = c_1 T^\alpha B^\beta$, and Figure~\ref{figure-6} to the mixed power-law model $f(T, B) = c_2 T^\alpha + c_3 B^\beta$.

As shown in Figure~\ref{figure-5}, the power-law model exhibits dynamical behavior analogous to that of a Chaplygin gas model before asymptotically approaching a $\Lambda$CDM-like regime. In Figure~\ref{figure-6}, although the mixed power-law model also converges to a $\Lambda$CDM-like state in the late-time limit, the evolutionary paths in the $r$-$q$ and $r$-$s$ planes differ significantly between the two parameter sets. For $(\alpha, \beta) = (5, -1000)$, the trajectory remains within the region characteristic of the Chaplygin gas model before reaching the $\Lambda$CDM point $\{r = 1, s = 0\}$. In contrast, for $(\alpha, \beta) = (5, 2/3)$, the trajectory crosses the $\Lambda$CDM point, showing damped transitions between the Chaplygin gas and quintessence regimes. Additionally, the trajectory for $(5, 2/3)$ displays a spiral structure in the $r$-$q$ plane.

In particular, the distinct evolutionary trajectories observed in the statefinder diagnostic planes highlight the sensitivity and discriminatory power of the $\left\{ r, q \right\}$ and $\left\{ r, s\right\} $ diagnostics in distinguishing between multiplicative power-law and additive mixed power-law forms of $f(T,B)$ models. The notable differences in both the paths and transitional behaviors between these two functional forms can be attributed to their distinct gravitational dynamics. The multiplicative form $f(T,B) = c_1 T^\alpha B^\beta$ introduces a strong and inseparable coupling between torsion and the boundary term, which constrains the evolutionary path to remain within Chaplygin gas-like regimes until late times. Conversely, the additive form $f(T,B) = c_2 T^\alpha + c_3 B^\beta$ decouples the contributions of $T$ and $B$, permitting more varied interactions. This decoupling enables transitions between Chaplygin gas and quintessence behaviors, and can even lead to oscillatory or spiral approaches to the $\Lambda$CDM attractor, as evident in the $r$-$q$ plane for the parameter set $\left( \alpha, \beta \right)$ = $\left( 5, 2/3 \right)$. Therefore, the statefinder analysis not only effectively distinguishes between the two $f(T,B)$ ansätzes but also reveals how the structural choice (namely, whether to couple or to separate $T$ and $B$) fundamentally shapes the dynamical character and potential transient phases of cosmic acceleration.

\section{Conclusions} \label{section-6}
This study systematically analyzes the cosmological dynamics of two well-motivated models within the framework of $f(T,B)$ modified gravity. Focusing on the multiplicative power-law form $f(T, B) = c_1 T^\alpha B^\beta$ and the additive mixed power-law form $f(T, B) = c_2 T^\alpha + c_3 B^\beta$, we examine how the coupling and decoupling of the torsion scalar $T$ and the boundary term $B$ shape the late-time evolution of a flat FLRW universe.\\
\indent
By constructing autonomous systems for both models, we identify stable de Sitter-type fixed points that act as late-time attractors, providing a purely geometric explanation for cosmic acceleration. The multiplicative power-law model shows a smooth convergence toward a $\Lambda$CDM-like state via an intermediate Chaplygin gas regime. In contrast, the additive mixed power-law model displays richer dynamical behavior, including damped oscillations and spiral trajectories in the statefinder planes, as illustrated for parameters such as $(\alpha, \beta) = (5, 2/3)$. Moreover, statefinder diagnostics, specifically the $r$-$s$ and $r$-$q$ planes, effectively distinguish each model from the other and from the standard $\Lambda$CDM scenario, highlighting observationally testable features.\\
\indent
Methodologically, our analysis differs from several earlier dynamical studies in modified gravity, such as those in \cite{Franco 2020, Samaddar 2023, Kadam 2023, Odintsov 2017}, which commonly adopt the parameterization $\lambda = \ddot{H}/H^3$. Instead, we introduce the auxiliary variables $y$ and $v$, which ensure the autonomy of the dynamical system without requiring additional phenomenological assumptions.\\
\indent
In comparison with $f(T) $ and $f(R) $ cosmologies, the present work highlights how the structural choice between multiplicative and additive coupling of $T $ and $B $ qualitatively influences the dynamical landscape. Unlike $f(T) $ or $f(R) $ models, the additive mixed power-law form of $f(T,B) $ permits richer transitional behaviors that are less common in simpler frameworks. These findings indicate that $f(T,B) $ gravity, especially in its additive form, provides a more flexible phenomenological framework for describing dynamical dark energy while naturally accounting for late-time cosmic acceleration.\\
\indent
Future research could naturally build on this foundation in several ways. Extending the dynamical analysis to explicitly include matter-dominated phases would provide a more complete description of cosmic evolution. The models should also be tested against a broader set of observational data, such as cosmic chronometers, baryon acoustic oscillations, and the growth of large-scale structure, to better constrain the parameters $\alpha $, $\beta $, and $c_i $. Furthermore, exploring more general functional forms of $f(T,B) $ (for example, logarithmic, exponential, or piecewise-defined combinations) could help assess the robustness of the dynamical features identified in this work. Another fruitful avenue would be to examine the implications of such models for early-universe cosmology, including scenarios related to inflation and singularity avoidance.

\section*{Acknowledgements}
The authors wish to express their sincere gratitude to the anonymous reviewers for their invaluable comments and constructive suggestions, which have significantly contributed to enhancing the quality and clarity of this work.

This work was supported by the National Natural Science Foundation of China (Grant No. 12172322), the Yangzhou Key Laboratory of Intelligent Data Processing and Security (Grant No. YZ2024245), the ``High-end Talent Support Program'' of Yangzhou University (2021), China, and the Postgraduate Research \& Practice Innovation Program of Jiangsu Province (Grant No. KYCX24\_3709), China.


\end{document}